\documentclass[seceq]{ptptex}

\usepackage{graphicx}
\usepackage{epsfig}
\usepackage{amssymb,amsmath}
\usepackage{pstricks}

\newcommand{\be}{\begin{equation}}
\newcommand{\ee}{\end{equation}}
\newcommand{\ba}{\begin{eqnarray}}
\newcommand{\ea}{\end{eqnarray}}
\newcommand{\bi}{\begin{itemize}}
\newcommand{\ei}{\end{itemize}}

\newcommand{\RR}{{\rm I\kern -.2em  R}} 
\newcommand{\eq}{Eq.~}

\newcommand{\fig}{Fig.~}

\def\lsi{\raise0.3ex\hbox{$<$\kern-0.75em\raise-1.1ex\hbox{$\sim$}}}
\def\gsi{\raise0.3ex\hbox{$>$\kern-0.75em\raise-1.1ex\hbox{$\sim$}}}
\newcommand{\lsim}{\mathop{\lsi}}
\newcommand{\gsim}{\mathop{\gsi}}

\title{%        %You can use \\ for explicit line-break
Status of Lattice Studies of the QCD Phase Diagram%
}

\author{%       %Use \scshape  for the family name
Owe \textsc{Philipsen}\footnote{in collaboration with Ph.~de Forcrand (ETH Z\"urich/CERN)}%
}

\inst{%     %Affiliation, neglected when [addenda] or [errata]
Institut f\"ur Theoretische Physik, Westf\"alische Wilhelms-Universit\"at M\"unster,
48149 M\"unster, Germany
}

\abst{%         %this abstract is neglected when [addenda] or [errata]
Determining the QCD phase diagram is a pressing task in view of its
relevance for nuclear and astro-particle physics programmes.
We review the current status of lattice calculations of the phase diagram in
the $(T,\mu_B)$-plane for baryon chemical potentials $\mu_B\lsim 500$ MeV. 
At $\mu_B=0$, simulations
of staggered fermion actions predict the quark hadron transition to be a crossover
in the continuum limit. As a baryon chemical potential is turned on, there is mounting 
evidence on coarse lattices for the crossover to weaken, rather than turning into
a true phase transition at a critical point, as predicted by earlier simulations. 
}

\begin{document}

\maketitle

\section{Introduction}

The QCD phase diagram has been the subject of intense research over the last ten years. 
Once fully determined, it will locate the regions of different forms of nuclear matter 
in the parameter space spanned by temperature $T$ and baryon chemical potential $\mu_B$.
Based on the fundamental property of asymptotic freedom, one expects at least 
three different regions: hadronic (low $\mu_B,T$), quark gluon plasma (high $T$) 
and colour-superconducting (high $\mu_B$, low $T$). For 
chemical potentials exceeding $\mu_B \gsim 1$ GeV, 
the situation may be more complicated with possible additional phases. \cite{wilc} \ 

Unfortunately, a quantitative calculation of the phase diagram from first principles is extraordinarily difficult. Since  QCD is strongly coupled on scales
$\lsim 1$ GeV, lattice simulations are the only tool to eventually give reliable answers, provided
that systematic errors are controlled. As we shall see, at present it is still a long way to
achieve this goal. In fact, lattice investigations at finite density are 
hampered by the ``sign problem'', and only approximate methods are available that work at small quark densities, $\mu=\mu_B/3\lsim T$. \cite{oprev,csrev} \ This adds further
systematic errors to those known
from zero density thermodynamics, like finite volume and discretisation effects. 
Accordingly, in this contribution we shall only consider the quark hadron transition at small densities. 
The widely accepted expectation is for 
a finite density first order phase transition
terminating in a critical endpoint, and an analytic crossover behaviour at $\mu=0$
(cf.~\fig\ref{tccomp} (right)). 
%Various models capturing the approximate centre and/or chiral symmetries of 
%QCD support this picture, and estimates for the critical endpoint have been 
%collected in \fig\ref{exp} (right). \cite{misha} \ 
%The scatter indicates that a first principles investigation is warranted.
% 
% \begin{figure}[t]
%\includegraphics[width=0.5\textwidth]{qm99fig2.eps}
%\includegraphics[width=0.5\textwidth]{tmudat.eps}
%\caption{\label{exp} 
%}
%\end{figure}
%

\begin{figure}[t]
%\vspace*{-0.3cm}
\includegraphics[width=0.25\textwidth]{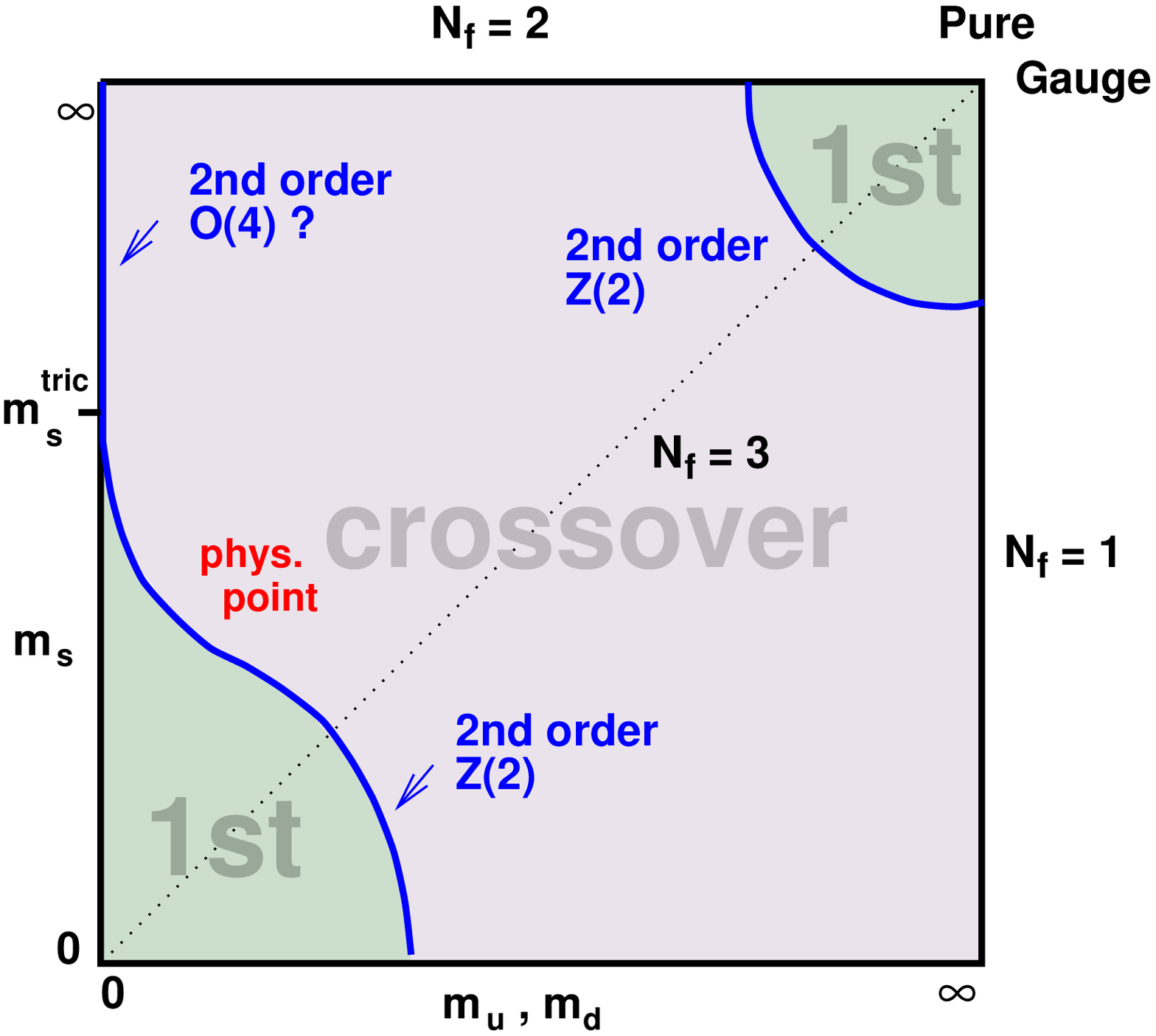}
\includegraphics[width=0.36\textwidth]{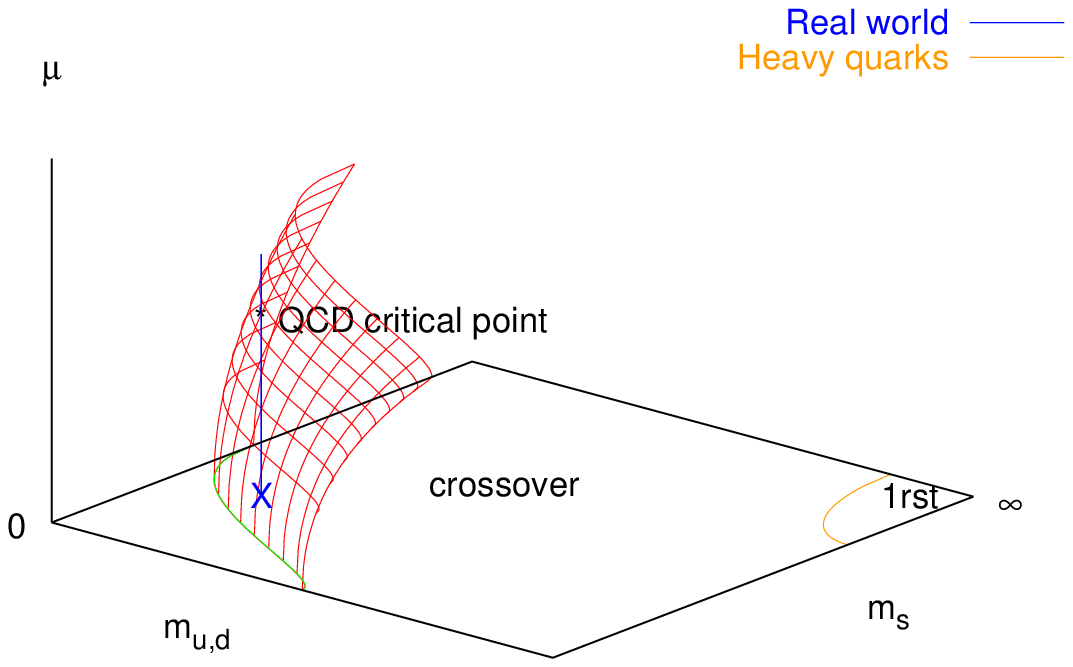}
\includegraphics[width=0.36\textwidth]{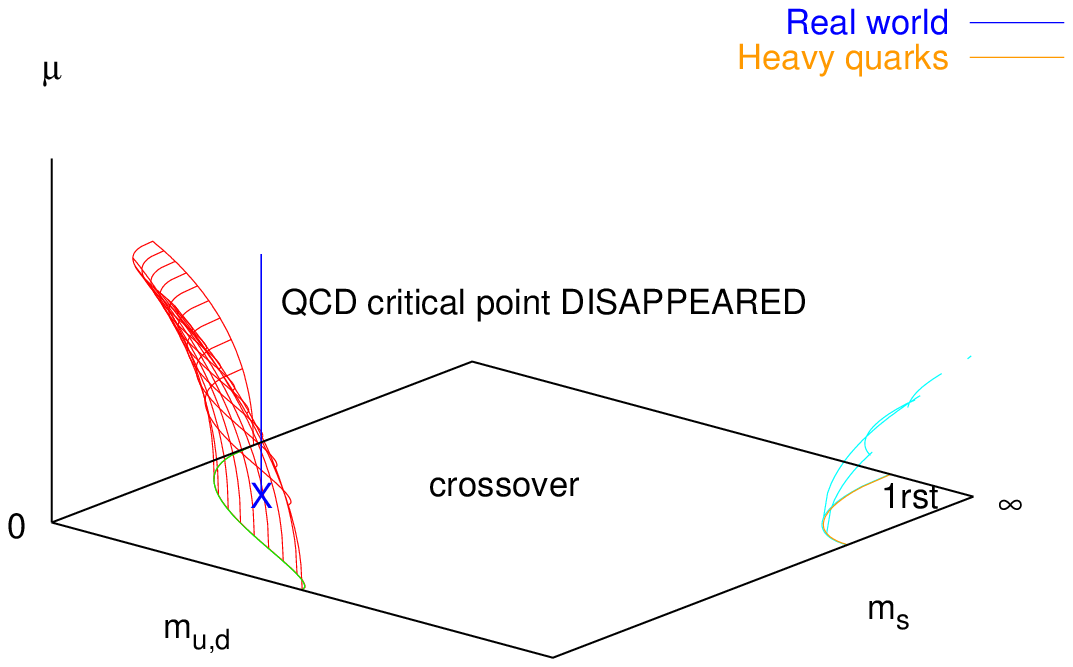}
\caption{\label{schem} Left: Schematic phase transition behaviour of $N_f=2+1$
QCD for different choices of quark masses $(m_{u,d},m_s)$ at
$\mu=0$. Middle/Right: Critical surface swept by the chiral 
critical line as $\mu$ is turned on. Depending on the curvature, a QCD chiral critical
point is present or absent. 
For heavy quarks the curvature has been determined \cite{kim1} and the first order 
region shrinks with $\mu$.
}
\end{figure}

In the absence of first principles calculations, where did this picture come from?
It is  based on
combining lattice results for $\mu=0$ in the larger parameter space 
$\{m_{u,d},m_s,T\}$ with model calculations at $\mu\neq 0$ \cite{misha} and 
connecting various limiting cases by universality and 
continuity arguments \cite{derivs}.
The schematic situation is depicted in \fig\ref{schem}.
In the limits of zero and infinite quark masses (lower left and upper 
right corners), order parameters corresponding to the breaking of a 
global symmetry can be defined, and one numerically finds first order phase
transitions at small and large quark masses at some finite
temperatures $T_c(m)$. On the other hand, one observes an analytic crossover at
intermediate quark masses, with second order boundary lines separating these
regions. Both lines have been shown to belong to the $Z(2)$ universality class
of the 3d Ising model \cite{kls,fp2,kim1}. 

%The chiral critical line
%has been mapped out on $N_t=4$ lattices, and the physical point is confirmed
%to be on the crossover side of this line \cite{deForcrand:2006pv,Aoki:2006we}.

The ``derivations'' of the generally expected QCD phase diagram  
\cite{derivs} state two crucial
assumptions: a) the chiral transition for $N_f=2$ is second order and thus in the
$O(4)$ universality class, which implies the existence of a tricritical point
at some strange quark mass $m_s^{tric}$; b) by switching on $\mu$, this point
will continuously move to larger $m_s$ until it ends up as a tricritical
point for the $N_f=2$ theory at some finite $\mu_{tric}$. Similarly,
for small but non-zero $m_{u,d}$, the $Z(2)$ chiral critical line would 
continously shift with $\mu$ until it passes through the physical point at $\mu_E$, corresponding to the endpoint of the QCD phase diagram. 
This is depicted in \fig\ref{schem} (middle), where the critical point is part of 
the chiral critical
surface. Note, however, that there is no a priori reason for this. In principle it is also
possible for the chiral critical surface to bend towards smaller quark masses, cf.~\fig\ref{schem} (right),
in which case there would be no chiral critical point or phase transition 
at moderate densities.
In the sequel the lattice evidence for these scenarios will be reviewed.

\section{$N_f=2$ at zero density}

Let us first consider assumption a) above. 
Since the cost of dynamical simulations explodes with 
shrinking quark masses, the order of the two-flavour chiral
transition is even harder to determine than that of physical QCD. Despite many attempts
it could not yet be conclusively settled.
Wilson fermions appear to see O(4) scaling \cite{wil}, 
while staggered actions are inconsistent with O(4) and O(2) (for the discretised theory) 
\cite{s2}. A recent finite size scaling analysis using staggered fermions
with unprecedented lattice sizes was performed in \cite{dig}. 
Again, these data appear inconsistent with O(4)/O(2), and the authors
conclude a first order transition to be more likely.
A different conclusion was reached in \cite{ksnf2}, in which $\chi$QCD was 
investigated numerically.
This is a staggered action modified by an irrelevant term 
(i.e.~one going to zero in the continuum limit) such as to allow 
simulations in the chiral limit. 
The authors find their data compatible with those of an 
$O(2)$ spin model on moderate to small volumes.
They thus suspect that finite volume effects of current $N_f=2$ QCD simulations
mask the correct scaling.

Finally, from universality of chiral models it is known that the order of the chiral transition
is related to the strength of the $U_A(1)$ anomaly \cite{piwi}.
In a model constructed to have the right symmetry with a tunable anomaly strength, it has recently been demonstrated non-perturbatively that both scenarios are possible,
with a strong anomaly required for the chiral phase transition to be second order \cite{chi}.
 
Thus, we cannot yet take assumption a) for granted. Should the chiral transition turn out 
to be first
order, the likely modification of  \fig\ref{schem} (left) 
would be the disappearance of the 
tricritical point, with the chiral critical line intersecting the $N_f=2$ axis at 
some finite $m_{u,d}$ and being Z(2)
all the way. Provided this line shifts to the right with $\mu$ as in 
assumption b), the $(T,\mu_B)$-phase diagram for physical QCD would
still look as expected. But contrary to the scenario in Ref.~\cite{derivs}, 
its critical point would be unrelated to any tricritical point. 

\section{The chiral critical line at $\mu=0$ \label{mu0}}

\begin{figure}[t]
 \hspace*{-1.5cm}
\includegraphics[width=0.5\textwidth]{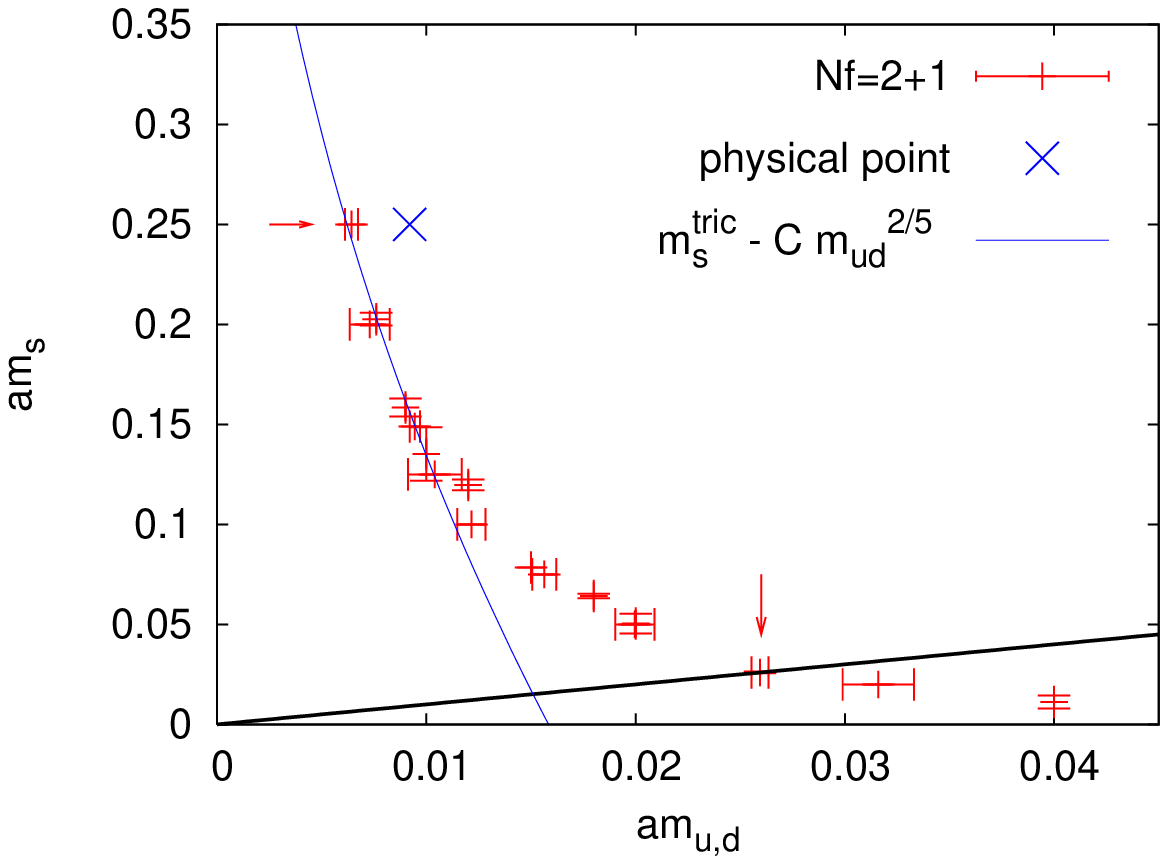}
\includegraphics*[bb=20 430 580 710,width=9.5cm]{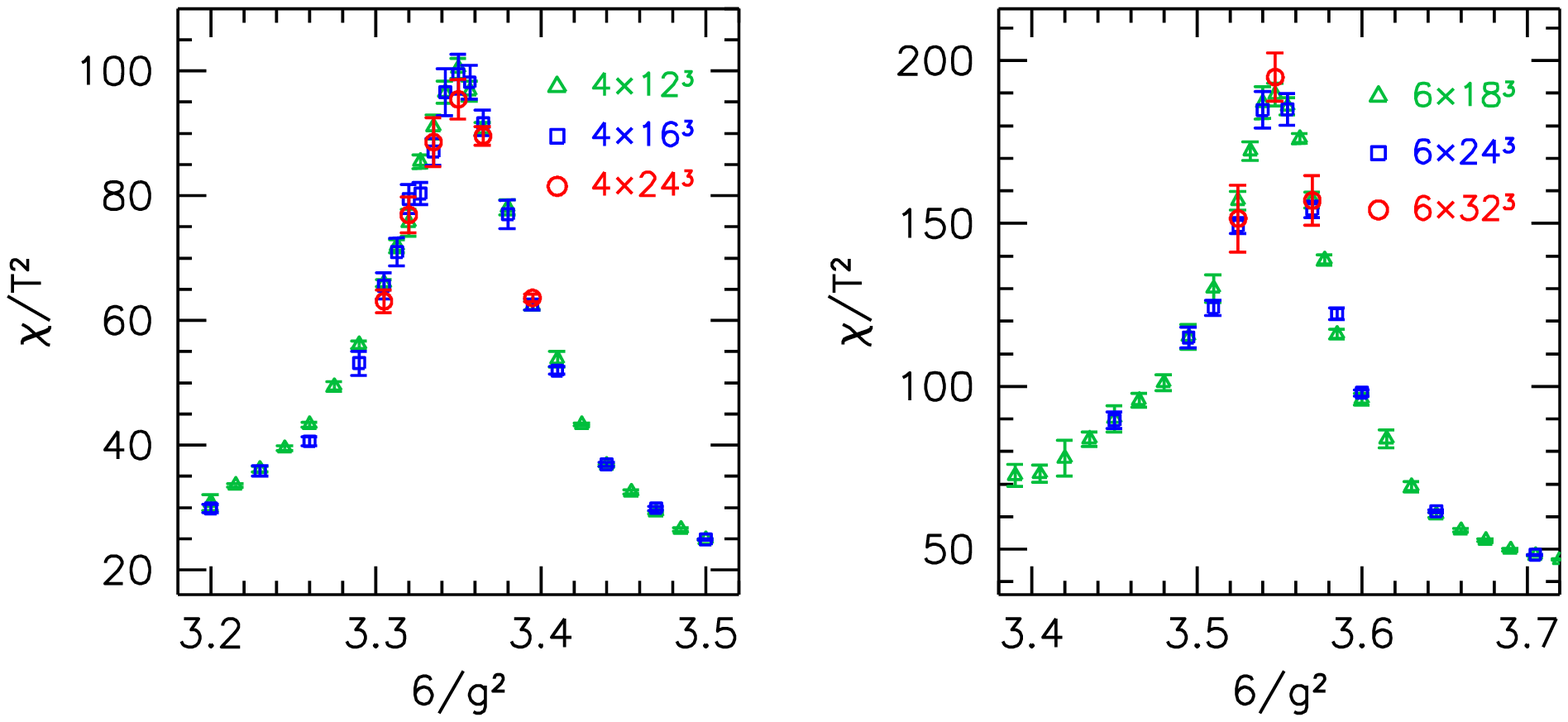}

\caption{\label{m1m2c} Left: The chiral critical line in the bare mass plane at $\mu=0$.
$N_f=3$ corresponds to the solid line. Also shown is a fit to an assumed tricritical point,
$m_s^{tric}\sim 2.8 T$ \cite{fp3}. Right: Finite size scaling of the chiral susceptibility
at the physical point for $N_t=4,6$. The peaks saturate at a finite value, consistent with
crossover behaviour. \cite{nature}
}
\end{figure}

The boundary line between the chiral first order and crossover regions has 
recently been mapped out on $N_t=4$ lattices
 \cite{fp3},  \fig\ref{m1m2c} (left).
A convenient observable is the Binder cumulant
$B_4(X) \equiv \langle (X - \langle X \rangle)^4 \rangle / \langle (X - \langle X \rangle)^2 \rangle^2$,
with $X \!=\! \bar\psi \psi$. At the second order transition, $B_4$ takes the value 1.604 dictated by the 
$3d$ Ising universality class.  
In agreement with expectations, the critical line steepens in approaching the chiral limit. Assuming a tricritical point on the $m_s$-axis according
to \fig\ref{schem} (left), the critical line is in fact consistent with tricritical scaling with $m_{u,d}$ 
\cite{derivs} and allows to estimate $m_s^{tric}\sim 2.8 T_c$. Note however, that this estimate is extremely cut-off sensitive and would change considerably on a finer lattice.      

The most important question concerns the location of the physical point, which is marked by the cross
in \fig\ref{m1m2c} (left). As expected, it is on the crossover side of the critical line. In Ref.~\citen{fp3} 
ratios of pion to rho and kaon masses were evaluated on the points marked by arrows,
to ensure that this statement indeed carries
over from bare quark masses to the spectrum of physical particles. Quark masses being extremely
susceptible to renormalisation effects, it is important to check this situation on finer lattices. This has been
completed in Ref.~\citen{nature} by a slightly different strategy. Here, the authors tune the quark masses
to the physical theory and then perform a finite size scaling analysis of susceptibilities around the critical
temperature, viz.~lattice coupling. This is shown in \fig\ref{m1m2c} (right) for different lattice spacings. Clearly, the peaks
saturate at a finite value which can be extrapolated to the continuum. This quite convincingly shows physical QCD to exhibit an analytic crossover rather than a true phase transition at zero density.
The only remaining caveat to this conclusion would be if there was a fundamental problem with the
so-called rooting trick when using staggered fermions, 
as frequently debated \cite{root}. Similar calculations using Wilson fermions 
could close this gap soon.

In an attempt to control cut-off effects, also 
the critical line has been checked on finer
$N_t\!=\!6$ ($a\sim 0.2$ fm) lattices.
The results show an important shift of the critical line
towards the origin: for the $N_f\!=\!3$ theory, the pion mass,
measured at $T\!=\!0$ with the critical quark mass, decreases from 
$1.6\; T_c$ to $0.95\; T_c$~\cite{LAT07}. Similar results are reported for 
$N_f=2+1$ \cite{fklat07}. Since cut-off effects on the physical point are much milder,
this considerably increases the distance of the critical 
surface to the physical point, \fig\ref{schem}. Regardless of the sign of the 
curvature, this trend alone makes a QCD chiral critical point at small $\mu/T \lsim 1$ 
less likely.

\begin{figure}[t!]
\vspace*{-1cm}
\begin{minipage}{0.5\textwidth}
\includegraphics[angle=-90,width=7cm]{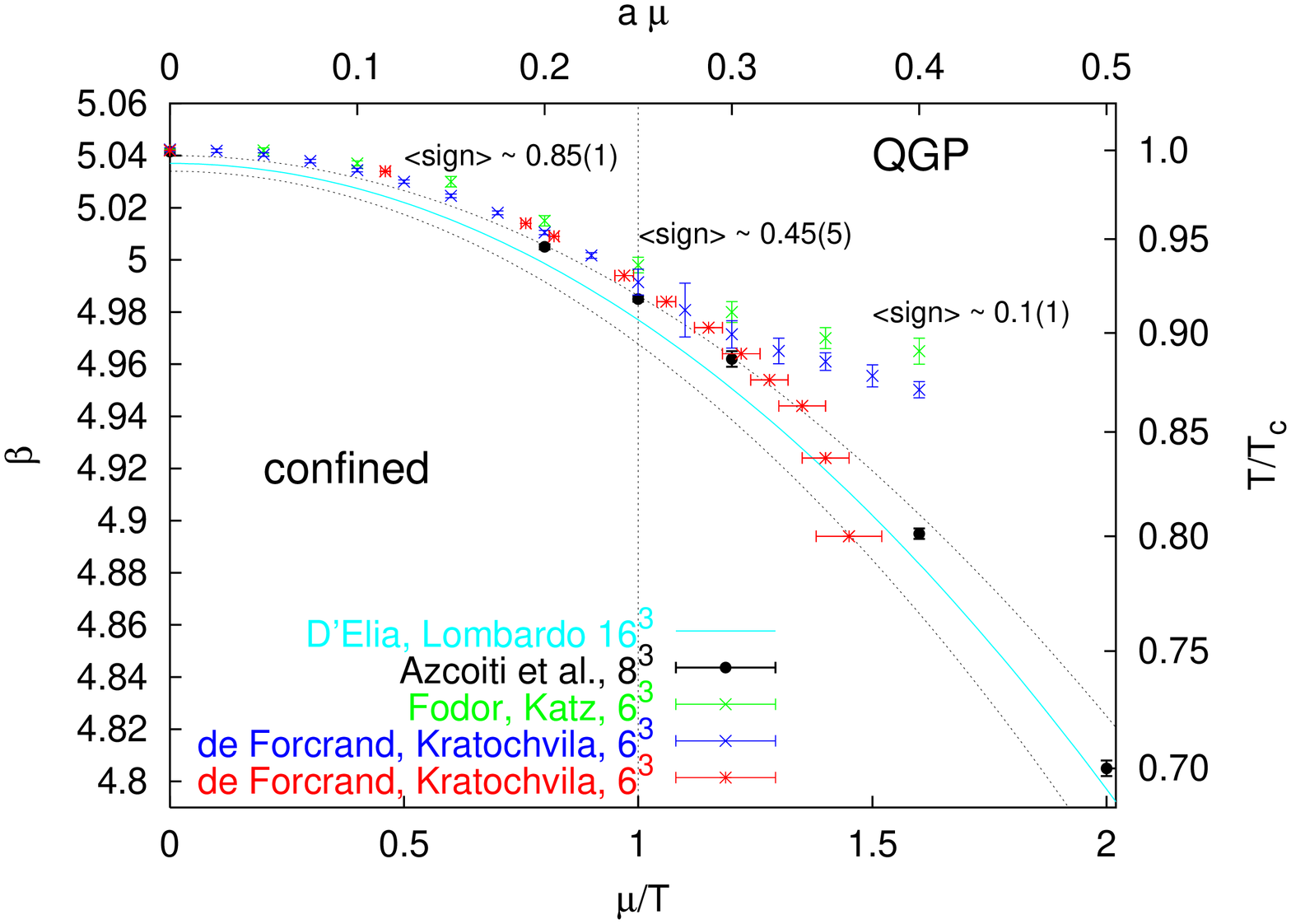}
%{\rotatebox{0}{\scalebox{0.4}{\includegraphics{phase_diag.eps}}}}
%\caption{Left: Comparison of different methods to compute the critical couplings. From\cite{}.}
\end{minipage}
\begin{minipage}{0.5\textwidth}
\includegraphics[width=7cm]{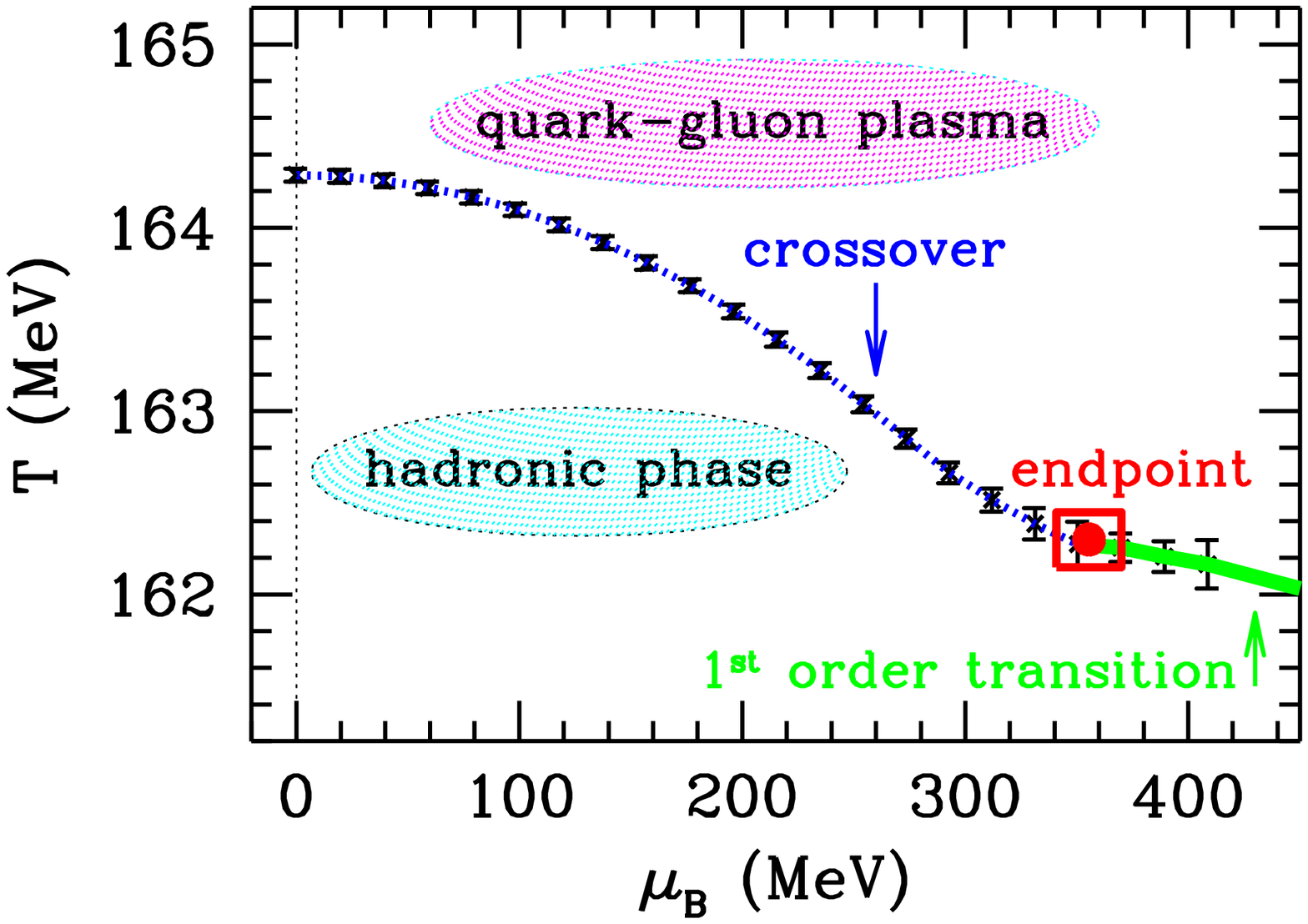}
\end{minipage}
\vspace*{-1cm}
\caption[]{Left: Comparison of different methods to compute the critical couplings
\cite{slavo}.
Right: Phase diagram for physical quark masses as predicted by two parameter 
reweighting on $N_t=4$ \cite{fk2}.}
\label{tccomp}
\end{figure}

\section{Calculations at finite density}

Straightforward Monte Carlo simulations at finite baryon density are impossible. 
This is because the
fermion determinant becomes complex for non-vanishing $\mu$,
prohibiting its use as a probability weight in Monte Carlo algorithms. 
This fact is also known as the 
``sign-problem''. 

There is a number of methods that circumvent the sign problem,
rather than solving it: i) Multi-parameter reweighting, 
ii) Taylor expansion in $(\mu/T)^2$ around $\mu=0$ (even powers because of CP invariance) 
and iii) simulations at imaginary 
chemical potential, either followed by analytic continuation or Fourier 
transformed to the canonical ensemble. All of these introduce some degree of 
approximation. However, the systematic errors are rather different, 
thus allowing for powerful crosschecks.  
%All methods are found to be reliable as long as $\mu/T\lsim 1$, or $\mu_B\lsim 550$ MeV, which includes the region of interest for heavy ion collisions. 
Reviews specialized on the technical aspects can be found in Refs.~\citen{oprev,csrev}. 

The first task is to identify the phase boundary, i.e.~the critical coupling and thus
$T_c(\mu)$.
This has been done for a variety of flavours and 
quark masses using different methods. For a quantitative comparison one 
needs data at one fixed parameter set.
Such a comparison is shown
for the critical coupling in \fig\ref{tccomp} (left), 
for $N_f=4$ staggered quarks with the same action and quark mass $m/T= 0.2$.
(For that quark mass the transition is first order
along the entire curve). One observes quantitative 
agreement up to $\mu/T\approx 1.3$, after which the different results start 
to scatter. Thus, all methods appear to be reliable for $\mu/T\lsim 1$, or $\mu_B\lsim 500$ MeV. 
% 
 %\begin{figure}[t]
 %\hspace*{-1cm}
%\includegraphics*[width=0.5\textwidth]{tc_nf.eps}
%\includegraphics[width=0.5\textwidth]{mu_nf2.eps}
%\caption{\label{b4mu} 
%}
%\end{figure}
%
The case of physical quark masses, after conversion to continuum units, 
is shown in \fig\ref{tccomp} (right) \cite{fk2}.
One observes that $T_c$ is decreasing only very 
slowly with $\mu$. This is consistent with a description by a 
series in $(\mu/\pi T)^2$ with
coefficients of order one,
\begin{equation}
\frac{T_c(\mu)}{T_c(0)}=1-t_2(N_f,m_f)\left(\frac{\mu}{\pi T}\right)^2
+{\mathcal O}\left( \left( \frac{\mu}{\pi T}\right)^4 \right) \quad .
\label{texp}
\end{equation}   
The leading coefficients for various cases have been collected from the 
literature \cite{csrev} and are
reproduced in Table \ref{tab:tccomp}. The curvature grows with 
$N_f$, which is 
consistent with $\sim N_f/N_c$ behaviour found in large $N_c$ expansions \cite{toublan}. 
Subleading coefficients are emerging at present but not statistically significant yet. 
Note that continuum conversions relying on the two-loop beta function 
are certainly not reliable for these coarse lattices, while fits to 
non-perturbative beta functions tend to increase the curvature.
\begin{table}[t]
\begin{center}
\begin{tabular}{cccccccc}\hline \hline
$N_f$&$am$&$N_s$&$t_2$&Action&$\beta$-Function&Method&Reference    \\ \hline
2  &0.1   & 16      &0.69(35) &p4   &non-pert.   &Taylor+Rew.& \cite{bisw1} \\
   &0.025 & 6,8     &0.500(34)&stag.&2-loop pert.&Imag.      & \cite{fp1}\\
3  &0.1   & 16      &0.247(59)&p4   &non-pert.   &Taylor+Rew.& \cite{bisw1}\\
   &0.026 & 8,12,16 &0.667(6) &stag.&2-loop pert.&Imag.      & \cite{fp3}\\
   &0.005 & 16      &1.13(45) &p4   &non-pert.   &Taylor+Rew.& \cite{bisw1}\\
4  &0.05  & 16      &0.93(9)  &stag.&2-loop pert.&Imag.      & \cite{el1}\\ \hline
2+1&0.0092,0.25&6-12&0.284(9) &stag.&non-pert.   &Rew.       & \cite{fk2}\\
\hline \hline
\end{tabular}
\caption{Coefficient $t_2$ in the Taylor
expansion of the transition line, \eq(\ref{texp}) from
$N_t=4$. \label{tab:tccomp}}
\end{center}
\end{table}

\section{The chiral critical surface}

All methods mentioned here also give signals for criticality, but the comparison
is non-trivial because of different parameter sets.
A simulation using reweighting methods on $N_t=4$ lattices puts the critical point at 
$\mu_B^E\sim 360$ MeV \cite{fk2}, \fig\ref{tccomp} (right), supporting the standard 
expected scenario. 
Quark masses were tuned to give the ratios 
$m_{\pi}/m_{\rho}\approx 0.19, m_{\pi}/m_K\approx 0.27$, which are close to their physical values.
In principle the determination of a critical point is also possible 
via the Taylor expansion, where
a true phase transition will be signalled by a finite radius of convergence for the 
pressure series about $\mu=0$ as the volume is increased. 
A critical endpoint for the $N_f=2$ theory, based on this approach, 
was reported  in \cite{ggpd}
for bare quark mass $m/T_c=0.1$. Taking the measured first four coefficients for the asymptotic behaviour of the series, the estimate 
for the location of the critical point is $\mu^E_B/T_E=1.1\pm 0.2$ at 
$T_E/T_c(\mu=0)=0.95$. 

 \begin{figure}[t]
 %\hspace*{-1cm}
\includegraphics[width=0.5\textwidth]{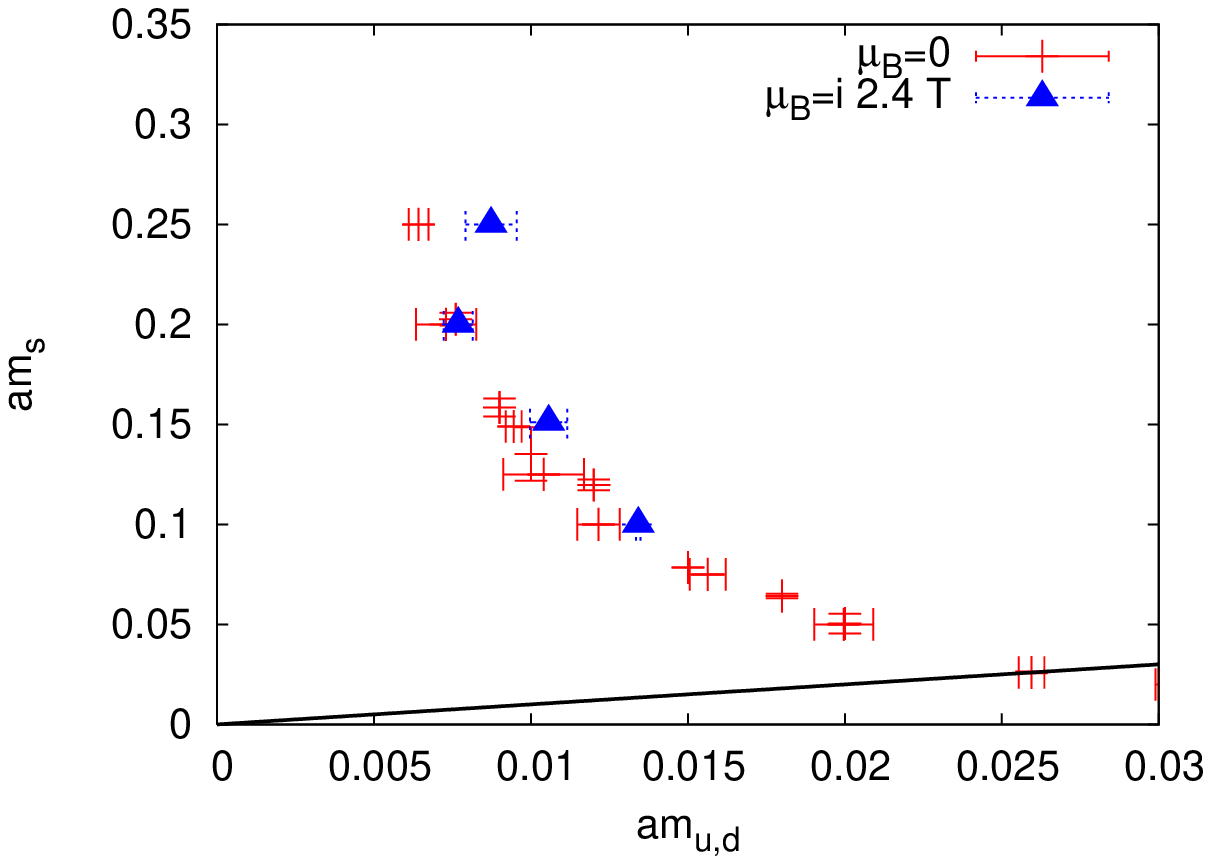}
\includegraphics[width=0.5\textwidth]{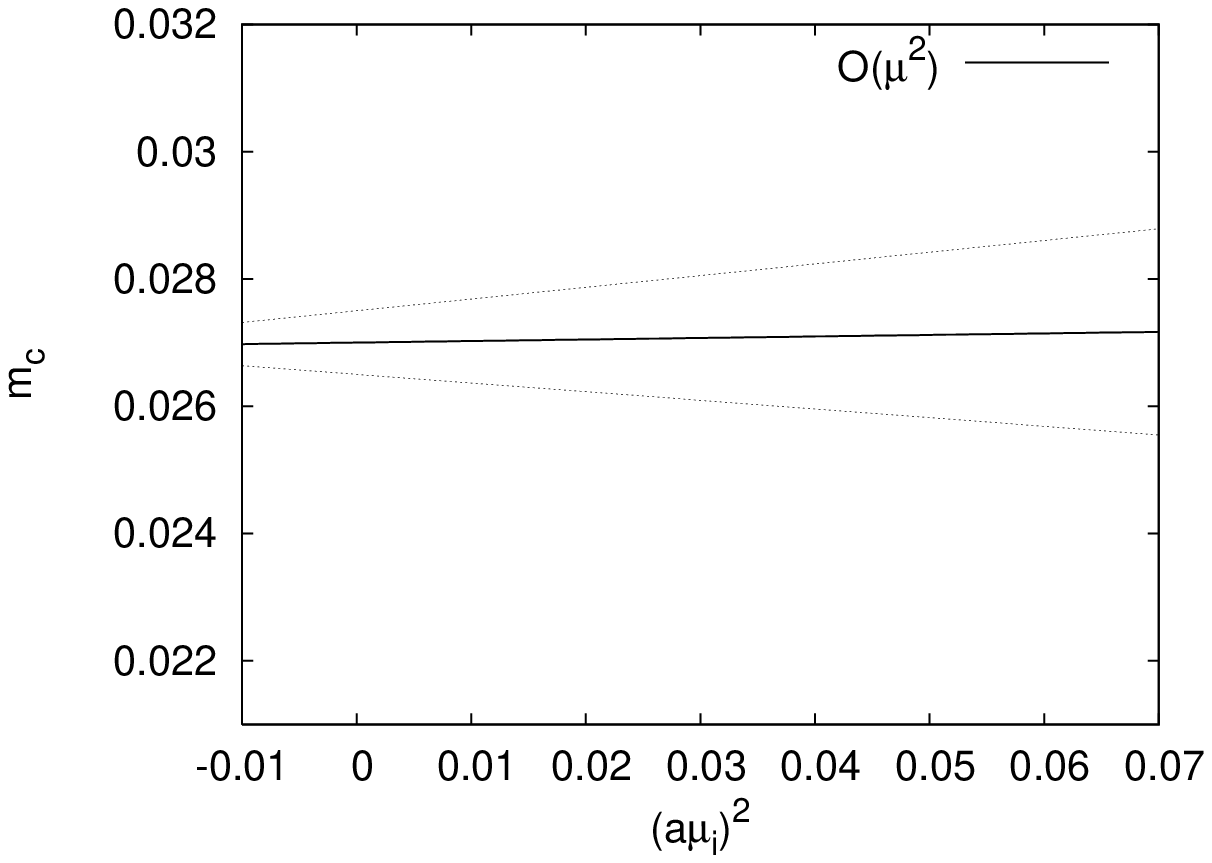}
\caption[]{\label{b4mu} Left: Chiral critical line at zero and non-zero imaginary $\mu$. 
Right: One sigma error band for the $N_f=3$ critical mass $am^c(\mu)$ resulting from a leading
order fit. Both for $N_t=4$ \cite{fp3}.
}
\end{figure}

Rather than fixing a theory with a particular set of quark masses and then switching on a chemical potential, let us now try to learn about the phase structure in the extended parameter space 
$\{m_{u,d},m_s,T,\mu\}$, i.e.~map out the chiral critical surface. This has been done for $N_t=4$ lattices using simulations at imaginary chemical potential \cite{fp3}. 
\fig\ref{b4mu} shows a comparison of the chiral critical line at zero density and a few points at imaginary 
chemical potential $\mu_B=i 2.4 T$. The finite density effect is very small, consistent with what is found for the change of the critical temperature, \eq(\ref{texp}). Thus, the critical surface in \fig\ref{schem} 
appears to emerge very steeply from the quark mass plane, making the critical point 
of physical QCD
extremely quark mass sensitive.  

The critical surface for imaginary $\mu$, \fig\ref{b4mu}, is moving to 
larger quark masses.
What does this imply for real chemical potential? To answer this question we focus on
$N_f=3$, collecting data for several values of $\mu_i$. Since 
$\mu_i/T\lsim 1$, the critical quark mass may be Taylor expanded 
$am^c(\mu)=am^c_0+c_1'(a\mu)^2+\ldots$, and the coefficients can be fitted to the 
data at imaginary $\mu$. 
\fig\ref{b4mu} (right) shows a one sigma error band for the critical bare quark mass 
from a leading order fit. As observed before,
the $\mu$-dependence is very weak
and even consistent with zero for these errors.
Since $T(\mu)=1/(a(\mu)N_t)$, $a(\mu)$ is an increasing function on a given lattice, so that
the critical mass in fixed physical units shrinks with $\mu$, as in \fig\ref{schem} (right).

\begin{figure}[t]
\includegraphics[width=0.5\textwidth]{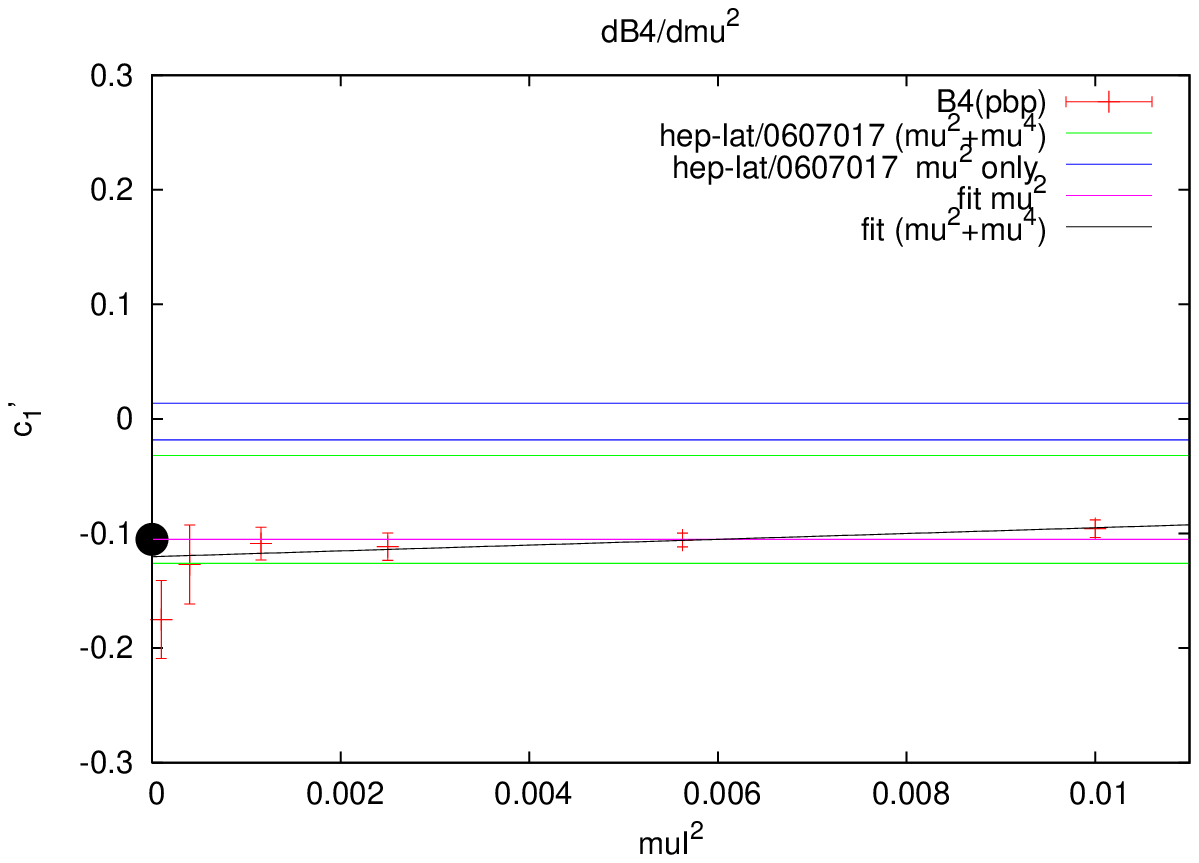}
\includegraphics[width=0.5\textwidth]{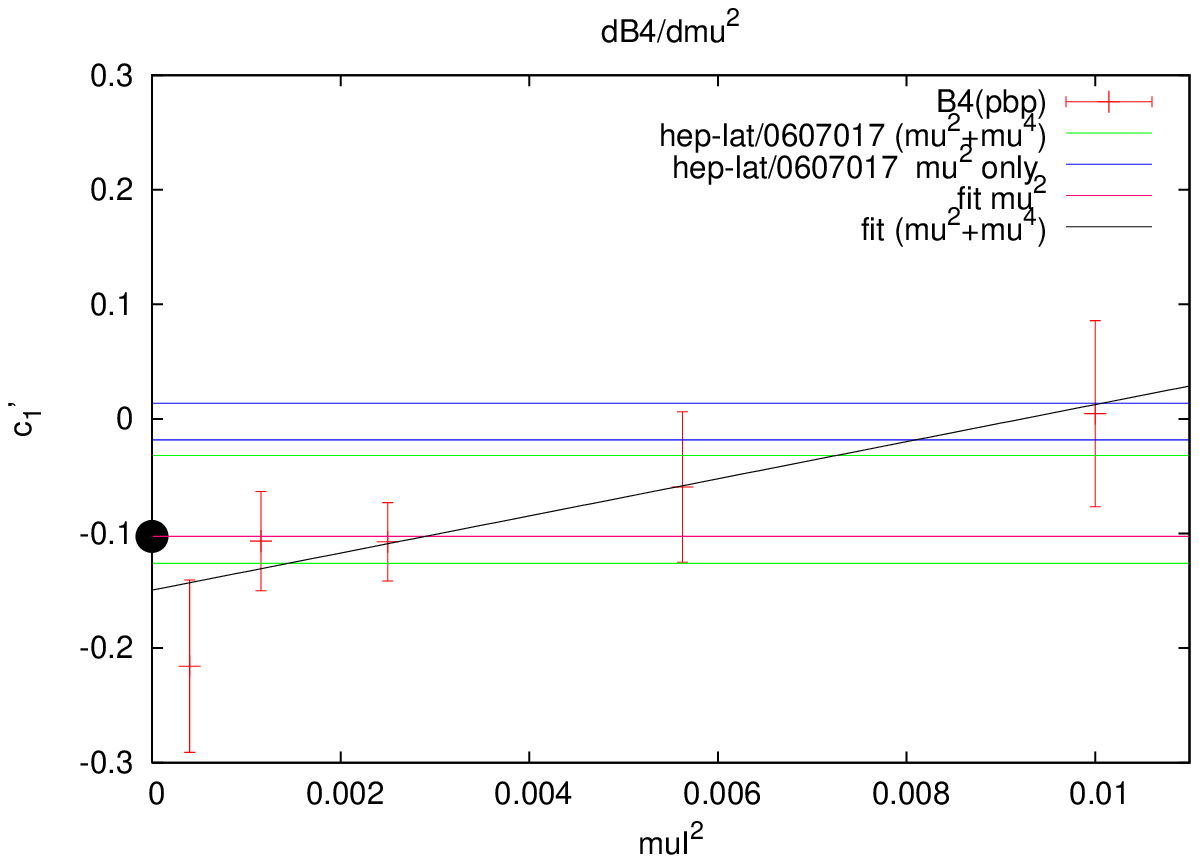}
\caption{\label{deriv} 
Leading $\mu^2$-derivatives of the critical quark mass on $N_t=4$ on $L=8$ (left) and
$L=12$ (right). The error bands give the corresponding values of LO and NLO fits
to Taylor series \cite{fp3}.
}
\end{figure}

One may worry about systematic errors when fitting a leading order polynomial 
to data containing
the full functional dependence. For example, subsequent terms may be 
cancelling in 
the imaginary, but not in the real direction. 
In order to check this, we have 
also calculated the leading derivative directly via 
$c_1'=-\partial_{(a\mu)^2}B_4/\partial_{am}B_4$. We do this in a novel, 
efficient way by evaluating finite differences
\be
\frac{\partial B_4}{\partial (a\mu)^2}=\lim_{(a\mu)^2\rightarrow 0}\frac{B_4(a\mu)-B_4(0)}{(a\mu)^2}.
\ee 
Because the required shift in the couplings is very small,
it is adequate and safe to use the original Monte Carlo ensemble 
for
$am^c_0,\mu=0$ and reweight the results by the standard 
Ferrenberg-Swendsen method. 
Moreover, by reweighting to imaginary $\mu$
the reweighting factors remain real positive and close to 1.
The results of this procedure 
based on 5 million trajectories on $8^3,12^3\times 4$, 
are shown in \fig\ref{deriv} for two volumes.
Subleading terms show up as a slope in the linear extrapolation. Indeed, such a slope
is visible in \fig\ref{deriv} (right). Both coefficients are
consistent with the results from the finite $\mu_i$ calculations, provided the 
next-to-leading order is taken into account.
We have also continued to collect statistics for the imaginary $\mu$ simulations, 
so we now have two significant terms. Putting everything together, 
our current best estimate
for $N_f=3$ on $N_t=4$ lattices is 
\be
\frac{m_c(\mu)}{m_c(0)} = 1 { - 3.3(5)} \left(\frac{\mu}{\pi T}\right)^2
{ - 12(6)} \left(\frac{\mu}{\pi T}\right)^4+\ldots
\label{final}
\ee
The emerging $\mu^4$-term is negative as well, further
shrinking the first order region with increasing $\mu$. On $N_t=4$ lattices, 
there is thus little doubt that the scenario \fig\ref{schem} (right) is realised 
for $\mu_B\lsim 500$ MeV. Note that this is analogous to the situation with heavy quarks,
where the first order transition is also weakening with $\mu$ \cite{kim1}.

\section{Discussion}

Our result for the chiral critical surface appears to be in contradiction with
the phase diagram obtained from reweighting methods, \fig\ref{tccomp}. As explained
above, we have checked
with independent methods that the signs we find for our coefficients are not artefacts of
the fitting procedure. On the other hand, there are concerns that the 
critical point determined via reweighting falls into a parameter range where reweighting 
becomes problematic \cite{kim}. However, even disregarding those, the bare lattice
results of Refs.~\citen{fp3,fk2} are not necessarily 
inconsistent. Reweighting in $\mu$ is performed
for fixed quark masses in lattice units, $am_q$. Since $T(\mu)=1/(aN_t)$, 
$a(\mu)$ is an increasing function on a given lattice, hence the critical point 
observed by reweighting corresponds to quark masses smaller than physical.
Indeed, \fig\ref{m1m2c} shows that on $N_t=4$ the physical point is very 
close to the critical surface. This is a discretisation effect, and simulations on 
finer lattices are required in order to settle the issue. As discussed above, 
we know already that the distance between the critical surface
and the physical point grows significantly, making a chiral critical point at small 
$\mu_B$ less likely, irrespective of the curvature of $m_c(\mu)$. Calculations of the 
curvature on $N_t=6$ are currently in progress.

%\section*{Acknowledgements}
%We would like to thank ...........

%\appendix
%\section{First Appendix} %Empty argument \section{} yields `Appendix'. 
%
%\section{Second Appendix}

\end{document}